\documentclass[aps,prx,twocolumn,showpacs,superscriptaddress,longbibliography]{revtex4-1}

\usepackage{amssymb}
\usepackage{amsfonts}
\usepackage{amsmath}
\usepackage{bm}
\usepackage{textcomp}
\usepackage{color}
\usepackage{longtable}
\usepackage{graphicx}
\usepackage{dcolumn}
\usepackage[a4paper=true,pagebackref=false]{hyperref}
\usepackage[normalem]{ulem}

\begin{document}

\title{Flipping coin experiment for studying switching in Josephson junctions and superconducting wires}

\author{M.~Foltyn}
\affiliation{Institute of Physics, Polish Academy of Sciences, Aleja Lotnikow 32/46, PL 02668 Warsaw, Poland}

\author{K.~Norowski}
\affiliation{Institute of Physics, Polish Academy of Sciences, Aleja Lotnikow 32/46, PL 02668 Warsaw, Poland}

\author{A.~Savin} \affiliation{Low Temperature Laboratory, Department of Applied Physics, Aalto University School of Science, P.O. Box 13500, 00076 Aalto, Finland}

\author{M.~Zgirski}
\email{zgirski@ifpan.edu.pl}

\affiliation{Institute of Physics, Polish Academy of Sciences, Aleja Lotnikow 32/46, PL 02668 Warsaw, Poland}

\date{\today}

\begin{abstract}
Josephson junctions and superconducting wires when probed with current pulses exhibit stochastic switching from superconducting to a stable non-zero voltage state. Electrical current dependence of the switching probability (so called S-curve) or switching current distribution is a fingerprint of the physics governing the escape process. This work addresses the criterion of independent switching event in a series of switching experiments. Treating Josephson junction as an electrical coin with current-tuned switching probability we investigate effect of correlation between switching events on the switching statistics.
\end{abstract}
\maketitle

Stochastic transition from superconducting to resistive state in Josephson junctions (JJ) and superconducting wires (SW) offers a workbench for studying decay of metastable states\cite{Fulton1974,Weiss2017,Clarke1985,Martinis1988,Hanggi1990} and opens doors to study quantum phenomena. The JJ switching phenomenon is used for probing the state of superconducting quantum bits\cite{Clarke2008, Chiorescu2003} and development of superconducting quantum information devices\cite{Choi2013}. Hysteretic JJ and SW are desired for threshold detection of various physical signals. They were employed for on-chip current measurements\cite{LeMasne2009}, studying thermal dynamics of nanostructures\cite{Zgirski2018} and proposed for single photon counting\cite{Walsh2017}.

It is common practice to measure electrical current dependence of the switching probability (so called S-curves)\cite{Pekola2005,Zgirski2011} or switching current distribution\cite{Krasnov2007,Bezryadin2009,Coskun2012,Kim2017} for the current biased JJ and SW. In the first case the sample is probed with train of $N$ current pulses and number of switchings $n$ yields estimation for the switching probability $p=n/N$. In the second case the sample is tested with current ramps and for each ramp switching current is recorded. Numbers of switchings in successive current intervals present switching current distribution. The shape of the measured S-curves or distributions reveals the information about a fundamental mechanism governing the transition from superconducting to normal state. The escape process is known to be driven either by thermal or quantum fluctuations\cite{Clarke1985,Pekola2005,Bezryadin2009,Lee2011}. In the former case the fitting Arrhenius-like relation to the experimental data allows to independently determine temperature of the electromagnetic environment. In the latter case the effective escape temperature is elevated above bath temperature and indicates the regime of the macroscopic quantum tunneling (MQT). Importantly, the temperature dependence of the S-curve width (or width of the switching current distribution) gives insight into the physical process responsible for the switching. For the orthodox thermally driven escape one observes monotonical increase in the S-curve width with temperature increase. The basic model of thermal activation for tunnel junctions yields the width $\Delta I \sim T^{2/3}$\cite{Weiss2017}. Intuitively it is understood as a thermal broadening, which is larger for higher temperatures. There are also cases when counterintuitive behavior is observed: reduction of the S-curve width when temperature is increased. There are a few phenomena responsible for such anticorrelation. In the moderately damped Josephson junction, as temperature is increased, the initial broadening of the switching threshold is followed by an apparent collapse of thermal activation\cite{Delsing2005, Pekola2005,Kim2017}. Such a reentrant behavior is attributed to retrapping process which sets the phase into diffusive motion and tends to keep the junction in the metastable state (so called phase diffusion regime). For the pure MQT the escape is governed by effective temperature defined as $T_{eff}=\hbar\omega_p/2\pi k_B$ with $\omega_p$ characterizing phase oscillations at the bottom of confining potential. For junctions $w_p$ scales with the critical current $I_C$. This corresponds to smaller widths of S-curve at higher temperatures for which $I_C$ is reduced. Narrowing of the observable switching current range with temperature was also demonstrated for superconducting wires\cite{Bezryadin2009, Goldbart2008}. It was attributed to multi-phase slip escape process understood as follows: Each phase slip is a dissipative event increasing temperature in the wire. For low temperature single phase slip is enough to heat the wire above $T_C$. At higher temperature a cascade of phase slips is required to exceed the $T_C$, with each phase slip increasing both temperature and probability for the next phase slip to occur.

\begin{figure}
\centering
\includegraphics[width=0.5\textwidth]{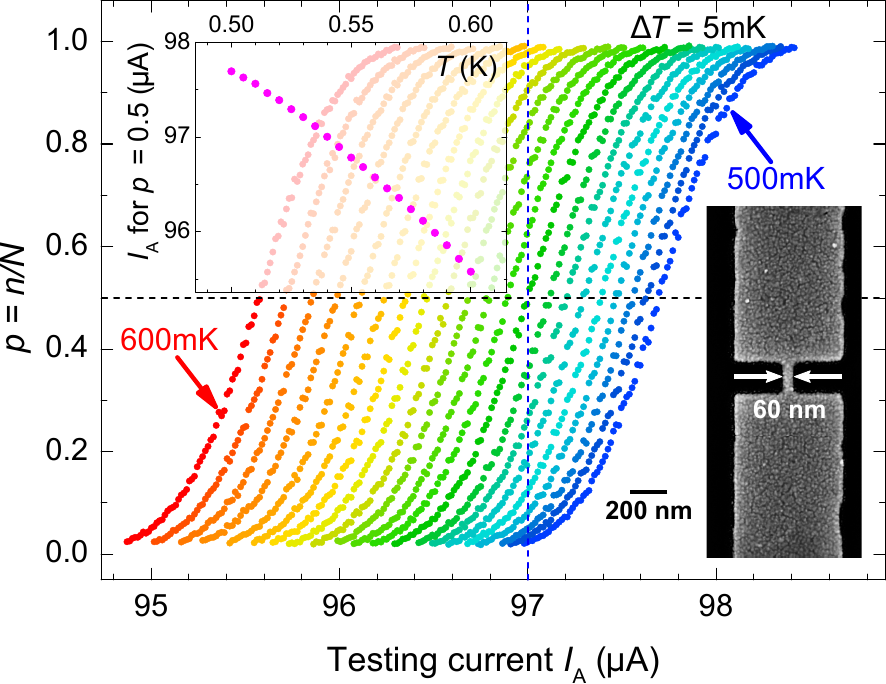}
\caption{\label{fig:Calib}S-curves measured at different bath temperatures used to extract switching current dependence on temperature (inset). Dashed lines correspond to $p$ = 0.5 (horizontal) and to a constant testing current (vertical). The SEM image of the aluminum nanobridge is presented on the right side.}
\end{figure}

\begin{figure}
\centering
\includegraphics[width=0.48\textwidth]{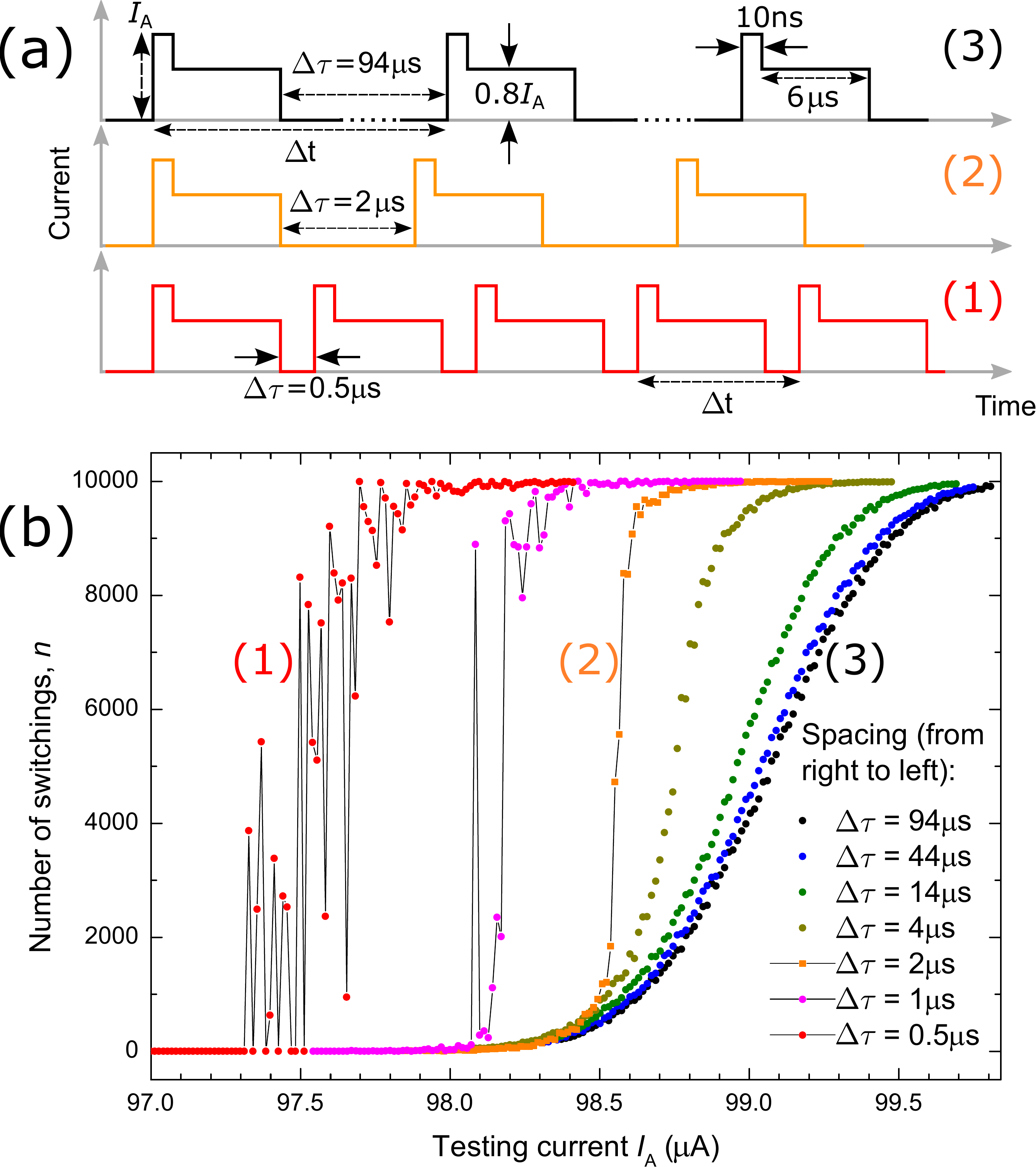}
\caption{\label{fig:Scurves}(a) Three trains of $N$ current pulses used to probe the bridge with different $\Delta \tau$ yielding dependencies (1), (2) and (3) shown in (b). (b) Current dependencies of the switching number $n$ for different spacing $\Delta \tau$ between probing pulses.}
\end{figure}

\begin{figure*}
\centering
\includegraphics[width=0.9\textwidth]{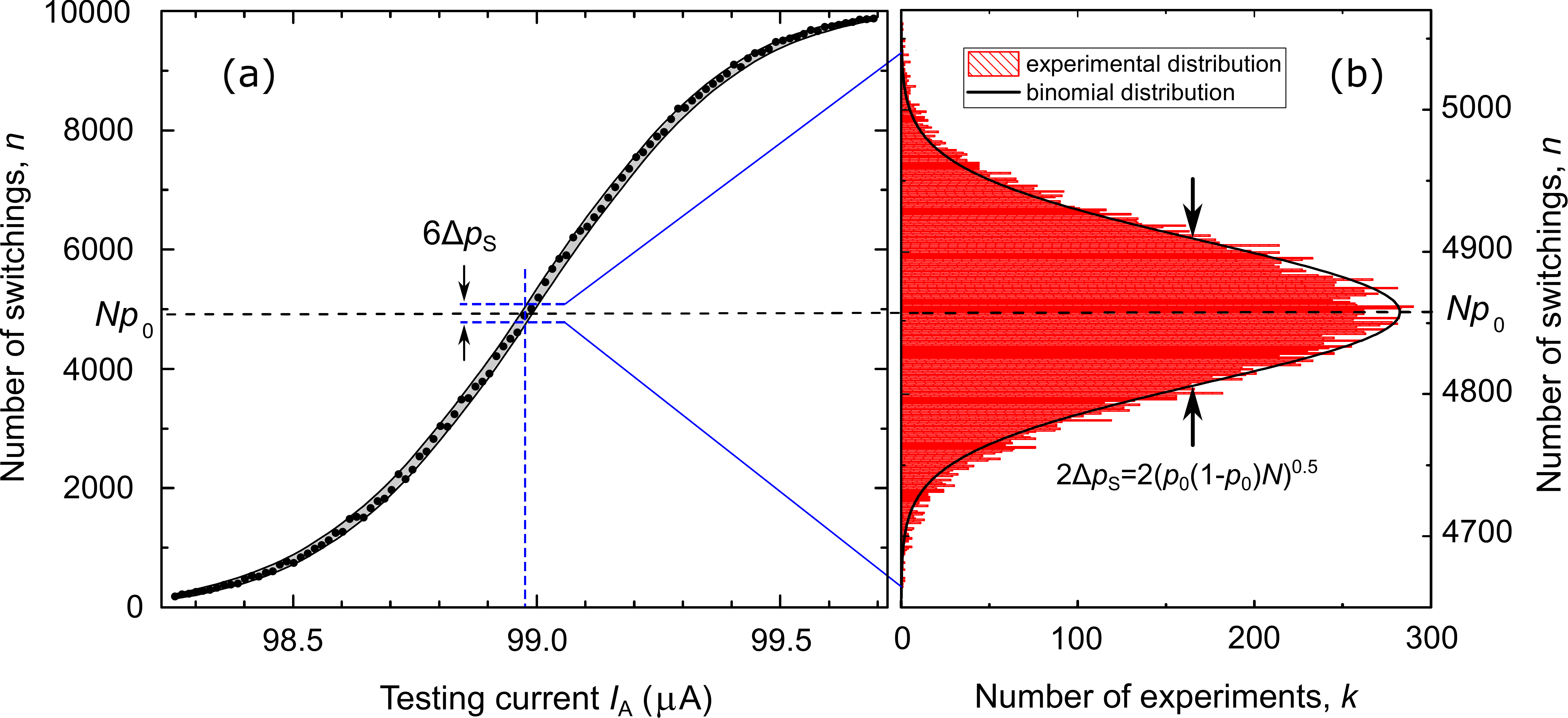}
\caption{\label{fig:Independent}"Flipping coin" experiment for the independent events at $T=300$mK. (a) The S-curve and its statistical broadening imposed as gray region. (b) Number of experiments (horizontal axis) resulting in the given number of switchings (vertical axis) for the constant current ($98.97\mu$A) indicated with dashed vertical line in (a). The single experiment consisted of sending $N=10000$ pulses and measuring the number of switching events. The experiment was repeated 35372 times. Binomial distribution is imposed with black curve. $\Delta p_S\cong50$ is the statistical broadening of the measurement at $p_0\cong$ 0.486.}
\end{figure*}

Probing of JJ and SW with current pulses has relied on a one silent assumption: JJ behaves like a coin for which "head and tail" experiment is performed. In the current work we answer the question: what makes a JJ a good coin? And what will happen if two successive "flips" are correlated?
A good coin should exhibit a stable value of probabilities of two possible outcomes (we relax here requirement that probability of each outcome should be 0.5). If this requirement is satisfied the expected number of outcomes of each kind in a series of many identical experiments, each consisting of a fixed number of flips, is given with binomial distribution.

On the other hand, if probability of getting the "head" is affected by result of previous flips we talk  about correlation. In the simplest case the correlation involves only two adjacent trials with earlier affecting later one (nearest neighbors correlation) but, as we will show below, it may have much more intricate character with result of each trial influenced by all previous outcomes. Switching of superconducting weak link gives us unique opportunity to realize and investigate both non-correlated and correlated switching scenarios.

We have fabricated superconducting nanobridge (width = 60nm) interrupting long nanowire (width = 600nm) connected to large area contact pads at both sides [Fig.\,\ref{fig:Calib}]. The structure was prepared by means of conventional one step e-beam lithography by evaporating of 30nm aluminum. We test the bridge with train of $N$ current pulses. In response to each pulse, dependently on the probing current amplitude and fluctuations, the bridge may remain in the superconducting state or transit to a normal state (switching). For low probing currents bridge never switches, for high current it always switches. In between there is a current region where bridge switching is probabilistic with probability rendering familiar S-shaped curve as the testing current increases [Fig.\,\ref{fig:Calib}]. For the detailed description of the method please refer to our earlier works\cite{Zgirski2015, Zgirski2018}. The switching probability for a fixed testing current increases with temperature (see vertical line in Fig.\,\ref{fig:Calib}). Further, we reserve notion of probability for independent events $p$, while generally we will talk about the switching number $n/N$ specifying number of switching events $n$ in the total number $N$ of probing pulses.

To investigate effect of correlation we intentionally introduce dependence between probing pulses using time interval between pulses $\Delta\tau$ as a knob for controlling the strength of correlation [Fig.\,\ref{fig:Scurves}(a)]. Current dependencies of the switching number $n(I_A)/N$ in the train of $N$ pulses with different separation time $\Delta\tau$ are presented in Fig.\,\ref{fig:Scurves}(b). For sufficiently large $\Delta\tau>50\mu$s the obtained S-curves are the same indicating that $\Delta\tau$ in this range does not influence the switching numbers. In such a case we talk about independent switching events and we can associate the switching number $n(I_A)/N$ with the independent switching probability $p$\cite{Flipping_coin_Note01}. With reduction of the period of the probing pulses, switching in a single pulse start to influence the result in subsequent pulses leading to steepening of the S-curve [Fig.\,\ref{fig:Scurves}(b)]. The further reduction of the period destroys familiar picture of S-curve: number of switchings $n(I_A)$ corresponding to a fixed probing current amplitude seems to be completely random as it is revealed on the scattered curves presented in Fig.\,\ref{fig:Scurves}(b). The observed correlation is of the thermal origin and appears when interval between probing pulses becomes shorter than thermal relaxation time for the bridge that switches to the normal state, thus exceeding $T_C$, and is left to cool down\cite{Zgirski2018}. The bridge does not reach the base temperature prior to arrival of the next testing pulse and switching probability is enhanced as it is shown in Fig.\,\ref{fig:Calib} for a fixed testing-current amplitude. The strength of correlation increases with reduction of the time interval between testing pulses. In the current work we focus on the distributions of switching number for the fixed probing current amplitude in 3 regimes: 1 $-$ for independent, 2 $-$ correlated, and 3 $-$ fully correlated switching events.

\begin{figure*}
\centering
\includegraphics[width=1.0\textwidth]{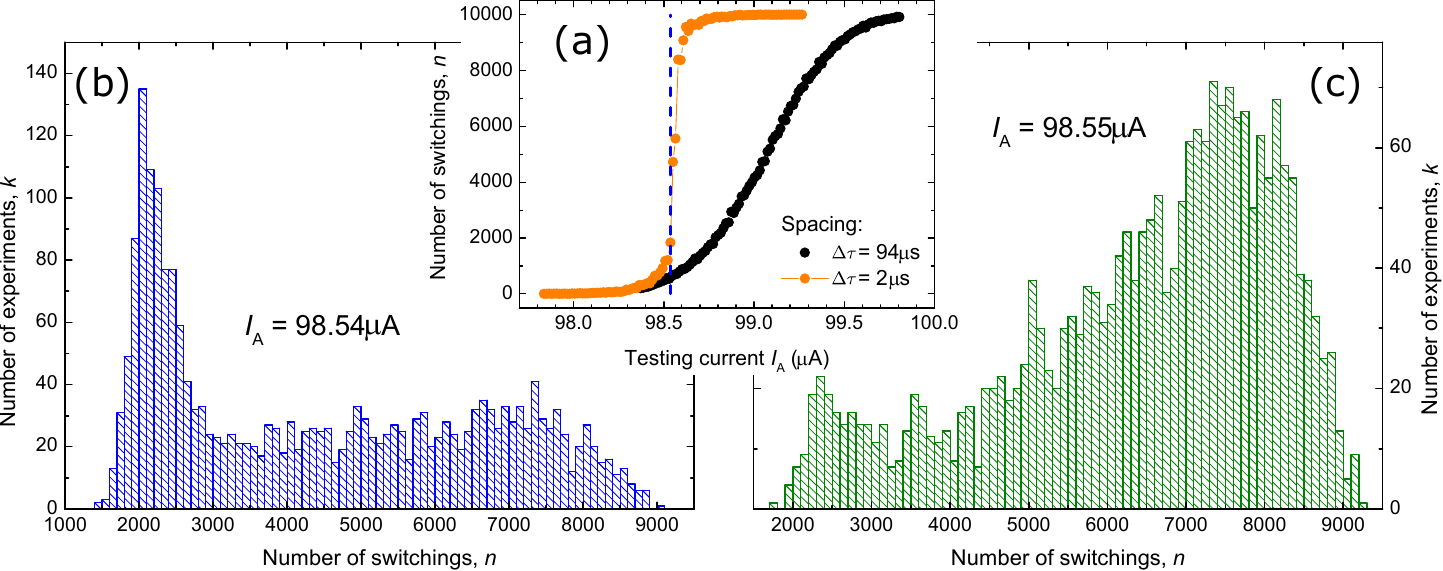}
\caption{\label{fig:Partially_correlated}"Flipping coin" experiment for the correlated events at $T=300$mK. (a) Two S-curves measured with pulse trains of different repetition rate (see Fig.\,\ref{fig:Scurves}). (b),(c) Number of experiments (vertical axis) resulting in the given number of switchings (horizontal axis) for the two slightly different current amplitudes indicated with dashed vertical line in (a) (single line shown for two currents). The single experiment consisted of sending $N=10000$ pulses and measuring the switching number. The experiment was repeated 2250 times.}
\end{figure*}

\begin{figure*}
\centering
\includegraphics[width=0.9\textwidth]{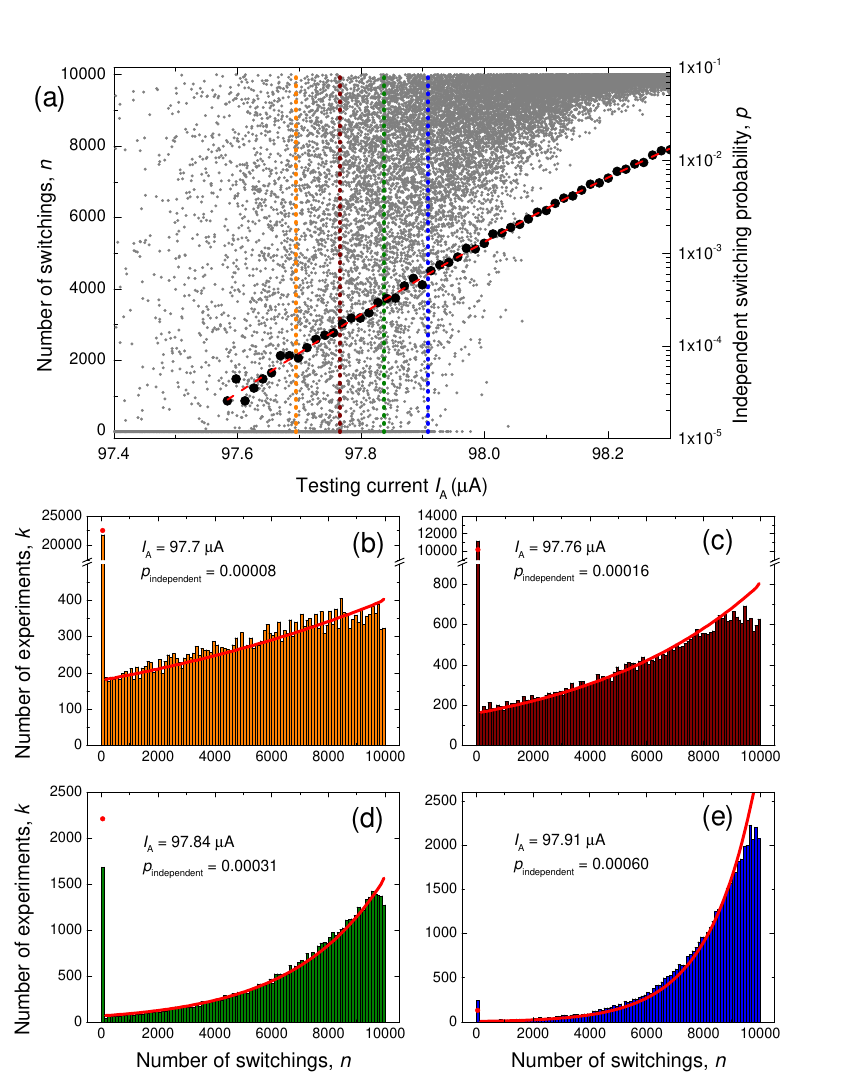}
\caption{\label{fig:Fully_correlated}"Flipping coin" experiment for the fully correlated events at $T=300$mK. (a) Random cloud of the switching numbers (left axis) for very slowly swept current amplitude with imposed independent switching probability measured in the limit of independent switching experiments (right axis). Each single point is the result of measurement with $N=10000$ pulses. (b),(c),(d),(e) Experimental distributions (number of experiments with corresponding number of switchings) recorded at constant current amplitudes indicated with dashed vertical lines in (a). The single experiment consisted of sending $N=10000$ pulses and measuring the switching number. The experiment was repeated 50000 times for each distribution. Red lines and dots are predictions of the panic distributions with experimentally determined independent switching probabilities $p$ presented in (a).}
\end{figure*}

We verify the assumption that independent switching events are described with binomial distribution:
\begin{equation}
P(n)={N \choose n}p^n (1-p)^{N-n}
\end{equation}
where $p$ is the independent switching probability and ${N \choose n}=N!/[(N-n)!n!]$ is number of different ways $n$ switchings can be distributed among $N$ trials. We send train of $N=10000$ current pulses of fixed amplitude with probing period of $\Delta t=100\mu$s and measure the switching number $n(I_A)/N$. Repeating the same experiment many times we reconstruct the switching distribution. Our procedure remains in the full analogy with flipping a coin (each pulse being a single toss) and indeed yields the binomial distribution [Fig.\,\ref{fig:Independent}], thus confirming independence of the switching events. The experimental distribution is a sensitive probe of the possible correlation between testing pulses. However, the distribution would be also affected if there was a temperature instability in the cryostat or excessive current noise leading to premature switching or preventing the bridge from switching. In such a case testing pulses may be not correlated, but still the distribution is violated. Nevertheless, if it is binomial with the proper variance, it serves as a strong indication of the independent switching events and negligible influence of electrical noise and temperature instabilities on the switching probability. The compliance with binomial distribution guarantees that measurements are only limited statistically: the measured probability exhibits the binomial broadening characteristic of finite number of trials. This broadening can be viewed as an unavoidable stochastic noise.

As we reduce the repetition time, the probing pulses become correlated. For $\Delta\tau= 2\mu$s (see Fig.\,\ref{fig:Scurves}a), when the S-curve becomes very steep, the resulting switching distributions are not binomial any more [Fig.\,\ref{fig:Partially_correlated}]. It is difficult to describe them by a compact, analytical distribution since involved correlations have long-range character with stochastic strength of correlation between pulses. The simplest numerical model, involving nearest pulses correlations only, could assume two values of the switching probability: $p$ for the case when there was no switching in the previous pulse and $q$ ($q>p$) if there was switching in the previous pulse. Such model has a very limited range of validity as it works only for the onset of correlations. Qualitatively, it is easy to observe, that it accounts for the steepening of S-curve. The observed steepening [Fig.\,\ref{fig:Scurves} and Fig.\,\ref{fig:Partially_correlated}] can be viewed as reduction of the apparent effective escape temperature $T_{eff}$ with $T_{eff}$ dependent on the strength of correlations. The case with $T_{eff}\approx 0$ is extremely sensitive to a change in the probing current. As far as S-curves can be measured in domain of current, temperature or magnetic flux, the S-curve with engineered effective temperature would be a desired building block for detectors e.g. it could be employed for sensing magnetization reversals that produce tiny changes in magnetic flux.

We now move on to the case of fully correlated pulses. We reduce the probing period to $\Delta t=6.5\mu$s ($\Delta \tau = $0.5$\mu$s, see Fig.\,\ref{fig:Scurves}a). Apparently, the switching number becomes fully unpredictable [Fig.\,\ref{fig:Fully_correlated}]. We enter regime when a single switching makes switching in the subsequent pulse certain leading to the switching avalanche. Such a phenomenon is described with the following switching number $n(I_A)$ probability distribution:
\begin{equation*}
P\left[n(I_A)\right]=(1-p)^{N-n}p
\end{equation*}
for $n(I_A) \ge 1$ and
\begin{equation*}
P\left[n(I_A)=0\right]=(1-p)^N
\end{equation*}
where $N$ is the number of probing pulses. Since all pulses before switching avalanche are not affected by previous testing pulses, $p$ is the independent switching probability measured in the independent switching regime [Fig.\,\ref{fig:Fully_correlated}(a) $-$ black dots].

We have coined the presented distribution \emph{the panic distribution}. In the Fig.\,\ref{fig:Fully_correlated}(b),(c),(d),(e) we present four experimentally measured distributions with the direct comparison to the postulated panic distribution. We have used $p$ measured in the independent switching regime [Fig.\,\ref{fig:Fully_correlated}(a) - black dots]. The remarkable agreement suggests that by performing measurements in the fully correlated regime one may measure independent switching probability when this probability is vanishingly small.

Interestingly, the panic distribution may describe social and economic variables in situations when a single person affects behavior of all other people from a certain group. It may also find application in situation when cascade of failures leads to breakdown of a system, for example electrical blackouts frequently result from cascade of failures between interdependent networks\cite{Buldyrev2010}.

We have proposed "the flipping coin" experiment as a tool for studying correlation in switching experiments. We have measured switching distributions for fixed current amplitudes in 3 regimes: for independent, correlated and fully correlated switching events. We have used the time interval between pulses as a knob for controlling correlations. Our experiment provides an interesting approach not only to study but also, perhaps more importantly, to engineer stochastic processes. We have shown that independent regime is manifested by the binomial distribution with the proper variance. We have demonstrated tuning of the apparent effective escape temperature in the correlated regime. We have found that in the fully correlated case the switching statistics is described with "the panic distribution" that exhibits high sensitivity to the independent switching probability. It is well known practice to measure magnetic field dependence of the switching current (i.e. Fraunhofer pattern) to prove the junction homogeneity prior to more advanced studies. Similarly, we propose to perform "the flipping coin" experiment to strengthen the quality of measured S-curves and switching current distributions, and, consequently, the credibility of the escape mechanism defining their shape.

The work is supported by Foundation for Polish Science (First TEAM/2016-1/10).


\begin{thebibliography}{23}
\bibitem [{Fli()}]{Flipping_coin_Note01}%
  \BibitemOpen
  \href@noop {} {}\bibinfo {note} {In fact what is measured in the independent
  switching regime is an estimator for the probability since measurement of the
  probability itself would need to involve infinite number of probing
  pulses.}\BibitemShut {Stop}%
\end{thebibliography}
\end{document}